\documentclass[10pt,aps,pra,showkeys,superscriptaddress]{revtex4-2}

\usepackage[colorlinks={true}]{hyperref}
\hypersetup{colorlinks=true,linkcolor=red,citecolor=blue,urlcolor=blue}
\usepackage{color,array}
\usepackage{amsmath, amssymb}
\usepackage{graphicx}
\usepackage{bm}
\usepackage{wrapfig}
\usepackage{subfigure}
\usepackage{epsfig}
\usepackage{epstopdf}
\usepackage{ar}
\usepackage{relsize}
\usepackage{dblfloatfix}
\usepackage{nicematrix}
\usepackage{orcidlink}
\usepackage{pstricks}
\usepackage{enumitem}
\usepackage{fixltx2e}
\usepackage{float}
\usepackage{placeins}
\usepackage{afterpage}

\begin{document}

\title{A comparative study of LQU and LQFI in general qubit-qutrit \\axially symmetric states}

\author{M.~A.~Yurischev\orcidlink{0000-0003-1719-3884}}\email{yur@itp.ac.ru}
\affiliation{
Federal Research Center of Problems of Chemical Physics and Medicinal Chemistry,
Russian Academy of Sciences, Chernogolovka 142432, Moscow Region, Russia
}

\author{Saeed~Haddadi\orcidlink{0000-0002-1596-0763}}\email{haddadi@semnan.ac.ir}
\affiliation{Faculty of Physics, Semnan University, P.O.Box 35195-363, Semnan, Iran}

\author{Mehrdad~Ghominejad\orcidlink{0000-0002-0136-7838}}\email{mghominejad@semnan.ac.ir}
\affiliation{Faculty of Physics, Semnan University, P.O.Box 35195-363, Semnan, Iran}

\begin{abstract}
\textbf{Abstract.} We derive the compact closed forms of local quantum uncertainty (LQU) and local quantum Fisher information (LQFI) for hybrid qubit-qutrit axially symmetric (AS) states.  This allows us to study the quantum correlations in detail and present some essentially novel results for spin-(1/2, 1) systems, the Hamiltonian of which contains ten independent types of physically important parameters. As an application of the derived formulas, we study the behavior of these two quantum correlation measures at thermal equilibrium. New features are observed in their behavior that are important for quantum information processing. Specifically, cascades of sudden changes in the behavior of LQU and LQFI are found with a smooth change in temperature or interaction parameters. Interestingly, in some cases, sudden transitions are observed in the behavior of LQU but not in LQFI, and vice versa.  Moreover,  our compact formulas open a way to apply them to other problems, for instance, when investigating the environmental effects on quantum correlations in open systems.
\end{abstract}

\keywords{
Qubit-qutrit system, axial $U(1)$ symmetry group, local quantum uncertainty, local quantum Fisher information
}

\maketitle
\section{Introduction}\label{sec:Intro}
Quantum correlations are at the heart of quantum information science, as they create unique capabilities that distinguish quantum systems from classical systems. These correlations are one of the distinguishing features of quantum mechanics and include phenomena such as quantum entanglement, where two or more particles are correlated in such a way that the state of one particle cannot be described independently of the state of the other particle(s). This implies that the properties of entangled particles are dependent on each other, regardless of the distance between them. For this reason, entanglement has emerged as a key resource in both quantum computing and quantum communication protocols \cite{Nielsen2000,Peres2002,Audretsch2007,McMahon2007} and so far many measures have been proposed to determine its value in bipartite and multipartite systems \cite{Plenio2007,Szalay2015,Saeedijtp2018}. To this end, some researchers are currently exploiting entanglement as a means to advance novel technologies that hold the promise of transforming the domains of computing, secure communication, and various other fields \cite{Amico2008,Horodecki2009,Gühne2009}.

While entanglement is the most well-known and studied form of quantum correlation, there exist other types of correlations beyond entanglement, such as quantum discord \cite{Zurk2000,Ollivier2001,Henderson2001,Streltsov2015}. It arises from the fact that quantum systems can have correlations that are purely non-classical in nature and cannot be attributed to classical information \cite{Ollivier2001}. In short, quantum discord is defined as the difference between two measures of total correlation, namely, mutual information and classical correlation \cite{Dakić2010}. There are several discord-like measures of quantum correlation that have been proposed in the literature \cite{dis1,dis2,dis3,dis4}, such as geometric quantum discord,  entanglement of formation-based quantum discord, measurement-induced disturbance, local quantum uncertainty (LQU) \cite{Girolami2013}, local quantum Fisher information (LQFI) \cite{Girolami2014}, and so forth. 

The measures LQU and LQFI have been introduced to capture different aspects of non-classical correlations between components of qubit-qudit ($2\otimes d$) systems. For instance, the authors of Ref. \cite{cap1} presented a comparative study of LQU and LQFI in the Heisenberg XY model. Khedif {\it et al.} \cite{cap2} considered a two-qubit Heisenberg XXZ model under an inhomogeneous magnetic field and investigated the thermal evolution of quantum correlations by means of concurrence, trace distance discord, and LQU. Besides, thermal LQU and LQFI in a two-qubit Heisenberg XYZ chain under Dzyaloshinsky–Moriya (DM) interaction were studied by Haseli \cite{cap3}. Also, QFI and skew information correlations in a pair of qubits coupled with dipolar and DM interactions at a thermal regime were investigated \cite{cap4}. Moreover, Fedorova and Yurischev \cite{cap5} considered a two-qubit Heisenberg chain with DM and Kaplan--Shekhtman--Entin-Wohlman--Aharony (KSEA) interactions at thermal equilibrium and examined the behavior of quantum discord, LQU,
and LQFI. Regarding the relationship between LQU and LQFI, Benabdallah {\it et al.} \cite{cap6} analyzed the quantum correlations under the presence of external magnetic field and intrinsic decoherence. 
Then, the behaviors of thermal non-classical correlations in two-coupled double quantum dots \cite{cap7} and spin square complexes \cite{cap8} were investigated by employing LQU and LQFI.
Recently, some authors explored LQU and LQFI in two gravitational cat states \cite{cap9}, an anisotropic two-qubit system \cite{cap10}, and a two-qubit-superconducting system \cite{cap11} (see also Refs. \cite{add1,add2,add3,add4,add5,add6} for further study).

A qubit-qudit system is a quantum state space with higher dimensions, thereby facilitating the execution of complex quantum operations and applications. For example, these hybrid systems can be used in quantum error-correction codes, where a qubit can represent logical information while a qudit can be employed for ancillary encoding, which provides additional error detection and error correction capabilities \cite{Chizzini2022}. Moreover, these systems enhance the security of quantum key distribution protocols, where qubits can be operated for secure key distribution while qudits can be employed for additional encoding or verification of the quantum channel \cite{Gisin2002}. Qubit-qudit systems can also be used in quantum communication protocols to transmit more complex quantum states than qubit-only systems, and interestingly, they can be employed in quantum state detection tasks, where the purpose is to distinguish between different quantum states. However, it is important to note that working with hybrid qubit-qudit systems may also be more technically demanding due to the increased complexity and the need for precise control over the quantum states involved. For this reason, further studies on hybrid quantum systems can lead to the engineering of different configurations that are more appropriate for quantum information processing tasks \cite{ad01,ad02,ad03,ad04,ad05,ad06,ad07,ad09}. 
For instance, some authors explored the thermal or time evolution of quantum correlations in hybrid qubit-qutrit systems such as spin chains \cite{S21,NBMR22,Tamer2023}, accelerated systems \cite{ad08,ad10,ad12}, and true-generalized/super-generalized X states \cite{ad13} under collective dephasing channel \cite{ad11}, random telegraph noise \cite{Benabdallah2022}, and intrinsic decoherence \cite{BARDAH23}.

In the recent paper \cite{YurischevPLA2023},  we provided explicit formulas of LQU  and LQFI for arbitrary two-qubit X states.
By extending this consideration, we derive here compact closed forms of LQU and LQFI for mixed qubit-qutrit ($2\otimes 3$) axially symmetric (AS) states.  These states play a crucial role in quantum information science and quantum mechanics due to their distinct mathematical properties and physical relevance. AS states exhibit rotational symmetry around a particular axis, which facilitates their mathematical representation and analysis. This symmetry reduces the number of independent parameters needed to describe the system and makes calculations more feasible. Many quantum tasks, such as quantum key distribution and certain forms of quantum teleportation, benefit from the structured nature of AS states. They often show enhanced coherence or entanglement properties under specific dynamics. Some molecular systems, including nanomagnets, cold atoms in optical traps, spin ensembles, superconducting qubits, or certain nuclear spin systems naturally exhibit AS properties.

The structure of our paper is as follows. 
In Sec. \ref{sec:Corr}, we obtain the explicit formulas of LQU and LQFI for the density matrix of a general qubit-qutrit AS state. As an example, we consider the described system at thermal equilibrium and discuss the results in Sec. \ref{sec:Gibbs}. Finally, our findings are summarized in Sec. \ref{subsubsec:Concl}.

\section{
Quantum correlations
}
\label{sec:Corr}

\subsection{
Local quantum uncertainty
}
\label{sec:LQU}
The concept of LQU refers to the intrinsic unpredictability of quantum phenomena at a particular location or in a local region in a quantum system.  It reflects the inherently probabilistic nature of quantum behavior and the limitations associated with the accuracy of simultaneous measurements of particular properties of quantum particles.

In order to quantify this concept, the LQU measure has been introduced (denoted by ${\cal U}$), which is derived from the Wigner-Yanase skew information $\mathcal{I}$ \cite{Wigner1963,Luo2003}.  When considering subsystem $A$ with the measurement operator $H_A$, and after optimization over all local observables on $A$, it is explicitly defined as \cite{Girolami2013}
\begin{equation}\label{lqu0}
\mathcal{U}(\rho)=\min _{H_A} \mathcal{I}\left(\rho, H_A\right).
\end{equation}

Specifically, the authors of Ref. \cite{Girolami2013} obtained the following expression for the qubit-qudit systems
\begin{equation}\label{lqu1}
{\cal U}=1-\lambda_{\max}^{(W)},
\end{equation}
where $\lambda_{\max}^{(W)}$ denotes the maximum eigenvalue of the  three-by-three symmetric
matrix $W$ whose entries are
\begin{equation}
   \label{eq:W}
   W_{\mu \nu}={\rm Tr}\{\rho^{1/2}(\sigma_\mu\otimes{\rm I_3})\rho^{1/2}(\sigma_\nu\otimes{\rm I_3})\},
\end{equation}
in which $\sigma_{\mu}$ and $\sigma_{\nu}$ are the set of Pauli matrices with $\mu, \nu=x, y, z$.

The diagonalization of the density matrix for a general qubit-qutrit AS state is presented in Methods \eqref{sec:Model}. Using Eqs.~(\ref{eq:xI3})--(\ref{eq:zI3}), we find that the matrix $W$ is
diagonal and its eigenvalues are equal to
\begin{eqnarray}
   \label{eq:Wxx}
   W_{xx}=W_{yy}=&&
   \frac{2}{(q_1^2+|u|^2)(q_2^2+|v|^2)}\big\{[\sqrt{p_1}(q_1^2\sqrt{p_3}+|u|^2\sqrt{p_2})(q_2^2+|v|^2)
	 +q_1^2\sqrt{p_2}(q_2^2\sqrt{p_5}+|v|^2\sqrt{p_4})
   \nonumber\\
   &&+\sqrt{p_4}[q_2^2\sqrt{p_6}(q_1^2+|u|^2)+|uv|^2\sqrt{p_3}]
	 +\sqrt{p_5}[q_2^2|u|^2\sqrt{p_3}+|v|^2(q_1^2+|u|^2)\sqrt{p_6}]\big\}
\end{eqnarray}
and
\begin{equation}
   \label{eq:Wzz}
   W_{zz}=p_1+p_6+\frac{8q_1^2|u|^2\sqrt{p_2p_3}+(p_2+p_3)(q_1^2-|u|^2)^2}{(q_1^2+|u|^2)^2}
	 +\frac{8q_2^2|v|^2\sqrt{p_4p_5}+(p_4+p_5)(q_2^2-|v|^2)^2}{(q_2^2+|v|^2)^2}.
\end{equation}
Due to $W_{xx}=W_{yy}$, this eigenvalue is two-fold degenerate.

Combining now manual and computer analytical calculations, we would be able to reduce the original Eqs.~(\ref{eq:Wxx}) and (\ref{eq:Wzz}) to the following compact
forms:
\begin{equation}
   \label{eq:Wxx1}
   W_{xx}=2\Biggl(\frac{c+\sqrt{p_2p_3}}{\sqrt{p_2}+\sqrt{p_3}}\sqrt{p_1}
   +\frac{b+\sqrt{p_4p_5}}{\sqrt{p_4}+\sqrt{p_5}}\sqrt{p_6}
   +\frac{a+\sqrt{p_2p_3}}{\sqrt{p_2}+\sqrt{p_3}}\cdot\frac{d+\sqrt{p_4p_5}}{\sqrt{p_4}+\sqrt{p_5}}\Biggr)
\end{equation}
and
\begin{equation}
   \label{eq:Wzz1}
   W_{zz}=1-(\sqrt{p_2}-\sqrt{p_3})^2-(\sqrt{p_4}-\sqrt{p_5})^2
	 +\frac{(a-c)^2}{(\sqrt{p_2}+\sqrt{p_3})^2}
	 +\frac{(b-d)^2}{(\sqrt{p_4}+\sqrt{p_5})^2}.
\end{equation}
As a result, the analytical formulas for the branches ${\cal U}_0=1-W_{zz}$ and
${\cal U}_1=1-W_{xx}$ take short compact forms 
\begin{equation}
   \label{eq:U0}
   {\cal U}_0=(\sqrt{p_2}-\sqrt{p_3})^2+(\sqrt{p_4}-\sqrt{p_5})^2
	 -\frac{(a-c)^2}{(\sqrt{p_2}+\sqrt{p_3})^2}
	 -\frac{(b-d)^2}{(\sqrt{p_4}+\sqrt{p_5})^2}
\end{equation}
and
\begin{equation}
   \label{eq:U1}
   {\cal U}_1=1-2\Biggl[\frac{c+\sqrt{p_2p_3}}{\sqrt{p_2}+\sqrt{p_3}}\sqrt{p_1}
   +\frac{b+\sqrt{p_4p_5}}{\sqrt{p_4}+\sqrt{p_5}}\sqrt{p_6}
   +\frac{(a+\sqrt{p_2p_3})(d+\sqrt{p_4p_5})}{(\sqrt{p_2}+\sqrt{p_3})(\sqrt{p_4}+\sqrt{p_5})}\Biggr].
\end{equation}
So, the formula of LQU for the general qubit-qutrit AS state is given by
\begin{equation}
   \label{eq:U}
   {\cal U}=\min\{{\cal U}_0, {\cal U}_1\}.
\end{equation}
Since the both quantities ${\cal U}_0$ and ${\cal U}_1$ are expressed
directly only via the matrix elements of the
density matrix (\ref{eq:rho}) and its eigenvalues (\ref{eq:p_2-5}),  hence, the same
would be valid for ${\cal U}$.

\subsection{
Local quantum Fisher information
}
\label{sec:LQFI}
The LQFI measure $\mathcal{F}$ based on QFI (denoted by $F$) is defined as follows \cite{Girolami2014}
\begin{equation}
\mathcal{F}(\rho)=\min _{H_A} F\left(\rho, H_A\right).
\end{equation}

In particular, if the subsystem $A$ is a qubit, the following formula can be written for the optimized LQFI as
\begin{equation}
{\cal F}=1-\lambda_{\max}^{(M)},
\end{equation}
where $\lambda_{\max}^{(M)}$ is the largest eigenvalue of the real symmetric three-by-three
matrix $M$ with entries
\begin{equation}
   \label{eq:M}
   M_{\mu\nu}=\sum_{\scriptstyle m,n\atop\scriptstyle p_m+p_n\ne0}\frac{2p_mp_n}{p_m+p_n}\langle m|\sigma_\mu\otimes I_3|n\rangle
	 \langle n|\sigma_\nu\otimes I_3|m\rangle.
\end{equation}

As for LQU, Eq. \eqref{lqu1}, the analytical evaluation of the quantum correlation LQFI for any qubit-qutrit state is possible in principle, but it requires solving a secular algebraic equation of the third degree using Cardano's formulas, which are, unfortunately, too complicated for practical use. On the other hand, although the AS reduces the number of free parameters in the model, it allows a solution only in square radicals.

Again using Eqs.~(\ref{eq:xI3})--(\ref{eq:zI3}),
we find that the matrix $M$ is also
diagonal and its eigenvalues are equal to
\begin{eqnarray}
   \label{eq:Mxx}
   M_{xx}=M_{yy}=&&\frac{4}{q_1^2+|u|^2}\Biggl(\frac{p_1p_3}{p_1+p_3}q_1^2+\frac{p_1p_2}{p_1+p_2}|u|^2\Biggr)
   +\frac{4}{q_2^2+|v|^2}\Biggl(\frac{p_4p_6}{p_4+p_6}q_2^2+\frac{p_5p_6}{p_5+p_6}|v|^2\Biggr)
   \nonumber\\
   &&+\frac{4}{(q_1^2+|u|^2)(q_2^2+|v|^2)}\Biggl(\frac{p_2p_4}{p_2+p_4}q_1^2|v|^2+\frac{p_2p_5}{p_2+p_5}q_1^2q_2^2
	 +\frac{p_3p_4}{p_3+p_4}|uv|^2+\frac{p_3p_5}{p_3+p_5}q_2^2|u|^2\Biggr)\nonumber\\
\end{eqnarray}
and
\begin{eqnarray}
   \label{eq:Mzz}
   M_{zz}=&&p_1+p_6+(p_2+p_3)\frac{(q_1^2-|u|^2)^2}{(q_1^2+|u|^2)^2}
   +\frac{16p_2p_3}{p_2+p_3}\frac{q_1^2|u|^2}{(q_1^2+|u|^2)^2}
   \nonumber\\
	 &&+(p_4+p_5)\frac{(q_2^2-|v|^2)^2}{(q_2^2+|v|^2)^2}+\frac{16p_4p_5}{p_4+p_5}\frac{q_2^2|v|^2}{(q_2^2+|v|^2)^2}.
\end{eqnarray}
Further handle and computer symbolic calculations lead to that
the branch ${\cal F}_0=1-M_{zz}$ is given as
\begin{equation}
   \label{eq:F0}
   {\cal F}_0=4\Biggl(\frac{|u|^2}{a+c}+\frac{|v|^2}{b+d}\Biggr)
\end{equation}
and
the branch ${\cal F}_1=1-M_{xx}$ is
\begin{eqnarray}
   \label{eq:F1}
   {\cal F}_1=&&1-4\Biggl[\frac{p_1(p_2p_3+p_1c)}{(p_1+p_2)(p_1+p_3)}+\frac{p_6(p_4p_5+p_6b)}{(p_4+p_6)(p_5+p_6)}\Biggr]-\frac{1}{(p_2-p_3)(p_4-p_5)}
   \nonumber\\
   &&\times\Bigg\{p_4(p_4-p_5-b+d)\Biggl[\frac{p_2(p_2-p_3+a-c)}{p_2+p_4}
   +\frac{p_3(p_2-p_3-a+c)}{p_3+p_4}\Biggr]
   \nonumber\\
   &&+p_5(p_4-p_5+b-d)\Biggl[\frac{p_2(p_2-p_3+a-c)}{p_2+p_5}
   +\frac{p_3(p_2-p_3-a+c)}{p_3+p_5}\Biggr]\Bigg\}.
\end{eqnarray}
Finally, the formula of LQFI for general qubit-qutrit AS state reads
\begin{equation}
   \label{eq:F}
   {\cal F}=\min\{{\cal F}_0,{\cal F}_1\}.
\end{equation}

So, we obtained compact closed  formulas (\ref{eq:U0}), (\ref{eq:U1}), (\ref{eq:F0})
and (\ref{eq:F1}) for branches ${\cal U}_0$, ${\cal U}_1$, ${\cal F}_0$ and
${\cal F}_1$.
By employing Eqs.~(\ref{eq:U}) and (\ref{eq:F}), one can easily calculate quantum
correlations LQU and LQFI for arbitrary qubit-qutrit AS density matrix (\ref{eq:rho})
directly only through its entries and eigenvalues.

\section{System at thermal equilibrium
}
\label{sec:Gibbs}
As an application of derived formulas for LQU and LQFI, we consider the behavior of
quantum correlations at thermal equilibrium in qubit-qutrit systems with axial
symmetry.

\subsection{
Hamiltonian and Gibbs density matrix
}
\label{subsec:Gibbs1}
The most general Hamiltonian of the qubit-qutrit system
commuting
with the ${\cal S}_z$ can be written as
\begin{eqnarray}
   \label{eq:H}
   {\cal H}
	 =&&B_1s_z+B_2S_z+J(s_xS_x+s_yS_y)+J_zs_zS_z+KS_z^2+K_1(S_x^2+S_y^2)+K_2s_zS_z^2
   \nonumber\\
&&+D_z(s_xS_y-s_yS_x)
+\Gamma[s_x(S_xS_z+S_zS_x)+s_y(S_yS_z+S_zS_y)]
   \nonumber\\
&&+\Lambda[s_x(S_yS_z+S_zS_y)-s_y(S_xS_z+S_zS_x)],
\end{eqnarray}
where $s_i=\sigma_i/2$ in which $\sigma_i$ are ordinary Pauli matrices,
and spin-1 matrices are given by
\begin{equation}
   \label{eq:S_xyz}
   S_x=\frac{1}{\sqrt2}
	 \left(
      \begin{array}{rrr}
      0&1&0\\
      1&0&1\\
			0&1&0
      \end{array}
   \right),\ 
   S_y=\frac{1}{\sqrt2}
	 \left(
      \begin{array}{rrr}
      0&-i&0\\
      i&0&-i\\
			0&i&0
      \end{array}
   \right),\ 
   S_z=
	 \left(
      \begin{array}{rrr}
      1&0&0\\
      0&0&0\\
			0&0&-1
      \end{array}
   \right).
\end{equation}
In Eq.~(\ref{eq:H}), the symbols of tensor product, $\otimes$, between spin-1/2 and spin-1
matrices,
as well as the identity matrices, $\rm I_2$ and $\rm I_3$, in the Zeeman terms are
omitted  
for the sake of simplicity.
The Hamiltonian (\ref{eq:H}) has ten parameters:
$B_1$ and $B_2$ are the $z$-components of an external magnetic field applied to spins
1/2 and 1, respectively;
$J$ and $J_z$ are the exchange Heisenberg constants;
$K$ and $K_1$ are the uniaxial and planar one-ion anisotropies, respectively;
$K_2$ is the uniaxial two-ion anisotropy;
$D_z$ is the $z$-component of Dzyaloshinsky vector;
$\Gamma$ and $\Lambda$ are new parameters introduced here that can be called respectively symmetric and asymmetric higher-order spin coupling terms.

Notably, this Hamiltonian contains many physically important cases. For example, $B_1$ and $B_2$ applied to spins 1/2 and 1, respectively, can affect the magnetic properties of compounds \cite{NBMR22,HNMTK99,Strecka2020}. 
The case
$J(s_xS_x+s_yS_y)+J_zs_zS_z$ is known as the Heisenberg XXZ mixed-spin (1/2, 1) model, however, its components can also be considered as a Hamiltonian to describe the hyperfine interaction in the case of deuterium,  
where $s_x$, $s_y$ and $s_z$ are the electron spin (spin-1/2) operators, and $S_x$, $S_y$ and $S_z$ are the
nuclear spin (spin-1) operators (see \cite{Wilmer1960,Karr2020} and references therein). Besides, its parts can be considered as a spin-orbit coupling term where $S_x$, $S_y$ and $S_z$ play a role in the orbital angular momentum operators.
Interestingly, the case ${\cal H}_{ss}= {\cal X} S_i^2 (i=x,y,z)$ can be interpreted as spin squeezing Hamiltonian under one-axis
twisting model where ${\cal X}~(\geq0)$ is the strength of the spin-squeezing interaction in $i$-direction \cite{KitagawaUeda}. The term DM interaction in the Hamiltonian ${\cal H}_{\textmd{DM}}=\textbf{D}\cdot (\textbf{s}_i \times \textbf{S}_j)$ is a type of anti-symmetric exchange interaction that occurs in magnetic materials without inversion symmetry. It plays a crucial role in the formation of chiral magnetic structures, spin Hall effects, magnetic phase transitions, spintronics applications, and so on.

\renewcommand{\thefootnote}{\arabic{footnote}}

In the matrix  
form, ${\cal H}$ has the AS structure 
\begin{equation}
   \label{eq:H1}
   {\cal H}=
	 \left(
      \begin{array}{cllccc}
      E_1&\ &\ &\ &\ \\
      \ &h_1&0&g_1&0&\ \\
			\ &0&h_2&0&g_2&\ \\
      \ &g_1^*&0&h_3&0&\ \\
      \ &0&g_2^*&0&h_4&\ \\
			\ &\ &\ &\ &\ &E_6
      \end{array}
   \right),
\end{equation}
where
\begin{eqnarray}
   \label{eq:h-g}
   &&h_1=B_1/2+2K_1,\qquad
   h_2=B_1/2-B_2-J_z/2+K+K_1+K_2/2,
   \nonumber\\
   &&h_3=-B_1/2+B_2-J_z/2+K+K_1-K_2/2,\qquad
   h_4=-B_1/2+2K_1,\\
   &&g_1=[J+\Gamma+i(D_z+\Lambda)]/\sqrt2,\qquad
   g_2=[J-\Gamma+i(D_z-\Lambda)]/\sqrt2.
   \nonumber
\end{eqnarray}
The energy levels are given by
\begin{eqnarray}
   \label{eq:Ei_}
   &&E_{1,6}=J_z/2+K+K_1\pm(B_1/2+B_2+K_2/2),
   \nonumber\\
   &&E_{2,3}=\frac{1}{2}(h_1+h_3\pm R_1)=\frac{1}{2}(B_2-J_z/2+K+3K_1-K_2/2\pm R_1),\\
   &&E_{4,5}=\frac{1}{2}(h_2+h_4\pm R_2)=\frac{1}{2}(-B_2-J_z/2+K+3K_1+K_2/2\pm R_2)
   \nonumber\\
\end{eqnarray}
with
\begin{equation}
   \label{eq:R1}
   R_1=\sqrt{(h_1-h_3)^2+4|g_1|^2}=\{(B_1-B_2+J_z/2-K+K_1+K_2/2)^2+2[(J+\Gamma)^2+(D_z+\Lambda)^2]\}^{1/2},
\end{equation}
\begin{equation}
   \label{eq:R2}
   R_2=\sqrt{(h_2-h_4)^2+4|g_2|^2}=\{(B_1-B_2-J_z/2-K-K_1+K_2/2)^2+2[(J-\Gamma)^2+(D_z-\Lambda)^2]\}^{1/2}.
\end{equation}
Note that ${\rm Tr}\,{\cal H}=4(K+2K_1)$.
To obtain Hamiltonian \eqref{eq:H}, we actually started with matrix \eqref{eq:H1}, expanded it into the Pauli and Gell-Mann matrices, and then used the relations of Gell-Mann matrices with spin-1 matrices \eqref{eq:S_xyz} and their powers; finally, having given physical meaning to the expansion coefficients, we came to the expression \eqref{eq:H}.

Finally, the Gibbs density matrix is defined as
\begin{equation}
   \label{eq:rhoG}
   \rho=\frac{1}{Z}\exp(-{\cal H}/T).
\end{equation}
Using now symbolic (analytical) calculations on a digital computer, we arrive at the Gibbs density matrix \eqref{eq:rho} with entries:

\begin{eqnarray}
   \label{eq:meG}
   &&p_1=\frac{1}{Z}e^{-E_1/T},
   \nonumber\\
   &&a=\frac{1}{Z}\Biggl(\cosh{\frac{R_1}{2T}}+\frac{h_3-h_1}{R_1}\sinh{\frac{R_1}{2T}}\Biggr)e^{-(h_1+h_3)/2T},
   \nonumber\\
   &&b=\frac{1}{Z}\Biggl(\cosh{\frac{R_2}{2T}}+\frac{h_4-h_2}{R_2}\sinh{\frac{R_2}{2T}}\Biggr)e^{-(h_2+h_4)/2T},
   \nonumber\\
   &&c=\frac{1}{Z}\Biggl(\cosh{\frac{R_1}{2T}}+\frac{h_1-h_3}{R_1}\sinh{\frac{R_1}{2T}}\Biggr)e^{-(h_1+h_3)/2T},
   \nonumber\\
   &&d=\frac{1}{Z}\Biggl(\cosh{\frac{R_2}{2T}}+\frac{h_2-h_4}{R_2}\sinh{\frac{R_2}{2T}}\Biggr)e^{-(h_2+h_4)/2T},
   \nonumber\\
   &&u=-\frac{2g_1}{ZR_1}\sinh{\frac{R_1}{2T}}e^{-(h_1+h_3)/2T},
   \nonumber\\
   &&v=-\frac{2g_2}{ZR_2}\sinh{\frac{R_2}{2T}}e^{-(h_2+h_4)/2T},
   \nonumber\\
   &&p_6=\frac{1}{Z}e^{-E_6/T}.
\end{eqnarray}
The partition function $Z=\sum_n\exp(-E_n/T)$ is 
expressed as
\begin{equation}
   \label{eq:Z}
   Z=2\Biggl[\cosh{\frac{B_1+2B_2+K_2}{2T}}e^{-(J_z+2K+2K_1)/2T}
   +\cosh{\frac{R_1}{2T}}e^{-(h_1+h_3)/2T}+\cosh{\frac{R_2}{2T}}e^{-(h_2+h_4)/2T}\Biggr].
\end{equation}

\subsection{
Behavior of quantum correlations
}
\label{subsec:UFvsT}

\subsubsection{
High-temperature behavior
}
\label{subsubsec:high-T}
Studying the behavior of thermal quantum correlations in physical systems at high temperatures is of a great importance. For this reason, here we examine the properties of LQU and LQFI at high temperatures.

From expression \eqref{eq:U0}, we obtain the first branch of LQU as
\begin{equation}
   \label{eq:U0_T8}
   {\cal U}_0(T)|_{T\to\infty}\approx\frac{J^2+D_z^2+\Gamma^2+\Lambda^2}{6T^2}+O(1/T^3)
\end{equation}
and the second branch of LQU, given in \eqref{eq:U1}, behaves as
\begin{equation}
   \label{eq:U1_T8}
   {\cal U}_1(T)|_{T\to\infty}\approx\frac{3B_1^2+4B_1K_2+2(J^2+J_z^2+K_2^2+D_z^2+\Gamma^2+\Lambda^2)}{24T^2}+O(1/T^3).
\end{equation}
Similarly, from expressions \eqref{eq:F0} and \eqref{eq:F1}, we arrive at the following formulas
\begin{equation}
   \label{eq:F0_T8}
   {\cal F}_0(T)|_{T\to\infty}\approx\frac{J^2+D_z^2+\Gamma^2+\Lambda^2}{3T^2}+O(1/T^3)
\end{equation}
and
\begin{equation}
   \label{eq:F1_T8}
   {\cal F}_1(T)|_{T\to\infty}\approx\frac{3B_1^2+4B_1K_2+2(J^2+J_z^2+K_2^2+D_z^2+\Gamma^2+\Lambda^2)}{12T^2}+O(1/T^3).
\end{equation}

Thus, the correlations drop according to the power law $T^{-2}$. Note that the main high-$T$ terms ${\cal F}_{0,1}$ are two times larger than the corresponding terms of ${\cal U}_{0,1}$.

\subsubsection{
Low-temperature limit
}
\label{subsubsec:low-T}
%
%
\begin{figure}[!t]
\begin{center}
\epsfig{file=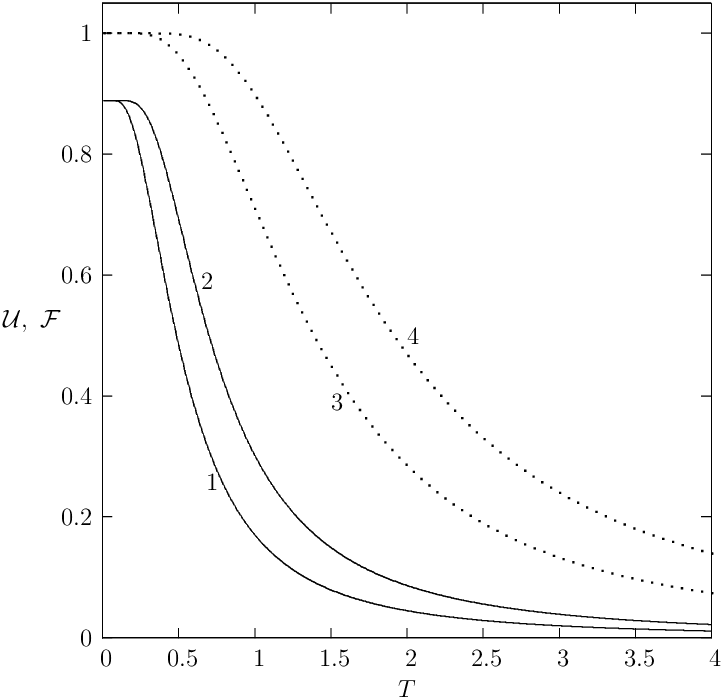,width=6cm}
\end{center}
\begin{center}
\caption{
Quantum correlations $\cal U$ (lines 1, 3) and $\cal F$ (lines 2, 4) of XXX systems
versus temperature $T$;
solid lines 1 and 2 correspond to spin-(1/2,1) model, while dotted lines 3 and 4
correspond to the two-qubit system with the Hamiltonian
${\cal H}=J{\vec\sigma}_1{\vec\sigma}_2$.
}
\label{fig:1}
\end{center}
\end{figure}
%
Let us consider the fully isotropic XXX model, i.e. put $J_z=J$ and all other eight parameters are equal to zero. The behavior of LQU and LQFI for the isotropic Heisenberg XXX (1/2,1) model
is shown in Fig.~\ref{fig:1}. For comparison, it is also shown the similar correlations for the two-qubit (1/2,1/2) system.

In the case under consideration, ${\cal U}_0={\cal U}_1$ and ${\cal F}_0={\cal F}_1$.
It is seen from Fig.~\ref{fig:1} that the correlations at $T=0$ are smaller for the qubit-qutrit system than for the two-qubit one.
Using Eq.~(\ref{eq:F0}), simple calculations give
\begin{equation}
   \label{eq:F0_XXX}
   {\cal F}_0(T)
	 =\frac{16}{9}\sinh{\frac{3J}{4T}}\tanh{\frac{3J}{4T}}\frac{e^{3J/4T}}{2+e^{3J/2T}}.
\end{equation}
So, we conclude that
\begin{equation}
   \label{eq:F0_0}
   \lim_{T\to0}{\cal F}(T)=8/9=0.888\ldots.
\end{equation}
The same is valid for LQU.
Remarkably, this limit does not depend on $J$ and therefore it is a universal value.

The analogous correlations in the two-qubit system are equal to one. Thus, the quantum correlation decreases with increasing spin $S$. This is due to that when the spin $S$ increases, the system is more classical and therefore quantum correlations decrease. A similar picture is observed for the entanglement \cite{S21}.

\subsubsection{Temperature dependence}
\label{subsubsec:sudden}
Remarkably, the behavior of quantum correlations can experience {\em abrupt} changes
under {\em smooth} varying of parameters of the system.

\begin{figure}[!t]
\begin{center}
\epsfig{file=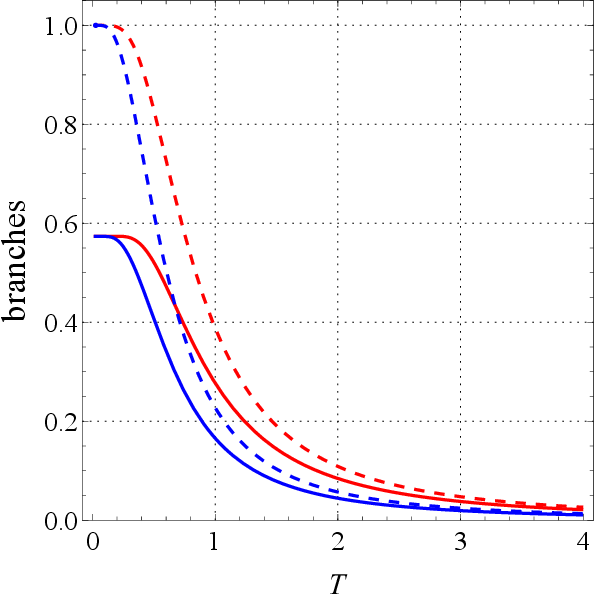,width=6cm}\put(-145,175){(a)}
\hspace{8mm}
\epsfig{file=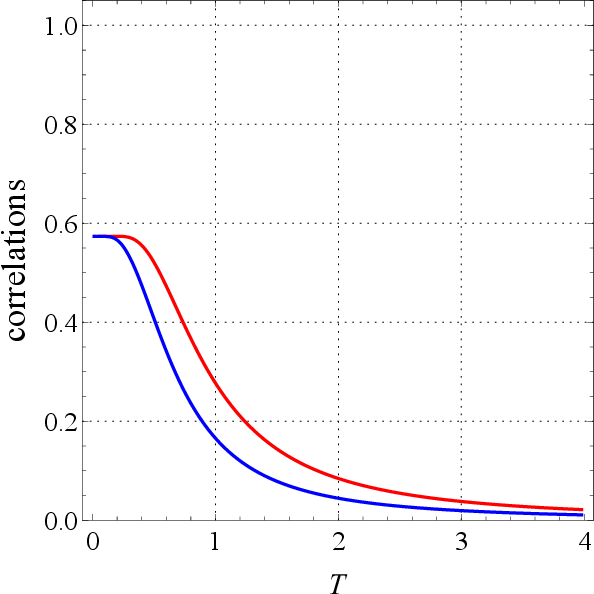,width=6cm}\put(-145,175){(b)}
\end{center}
\begin{center}
\caption{
Qubit-qutrit system with 
$B_1=0.3$, $B_2=-0.7$, $J=0$, $J_z=1$, $K=0.2$, $K_1=-0.1$, $K_2=0.22$, $D_z=0.32$,
$\Gamma=-0.87$, and $\Lambda=0.31$.
(a)~Branches of quantum correlations
${\cal U}_0$ (blue solid line),
${\cal U}_1$ (blue dashed line),
${\cal F}_0$ (red solid line), and
${\cal F}_1$ (red dashed line)
as a function of $T$.
(b)~Quantum correlations $\cal U$ (blue line) and $\cal F$ (red line) versus  $T$. 
}
\label{fig:2}
\end{center}
\end{figure}

Consider for example a random case as
$B_1=0.3$, $B_2=-0.7$, $J=0$, $J_z=1$, $K=0.2$, $K_1=-0.1$, $K_2=0.22$, $D_z=0.32$,
$\Gamma=-0.87$, and $\Lambda=0.31$. As can be seen from Fig. \ref{fig:2}, both correlations ${\cal U}=\min\{{\cal U}_0,{\cal U}_1\}$ (blue line)
and ${\cal F}=\min\{{\cal F}_0,{\cal F}_1\}$ (red line) are only determined by the branches ${\cal U}_0$ and ${\cal F}_0$, and no sharp changes are observed in the behaviors of LQU and LQFI at all temperatures. As a result, the correlations diminish monotonically from $0.57350$ to $0$ when the temperature $T$ goes from zero to infinity. 

\begin{figure}[!t]
\begin{center}
\epsfig{file=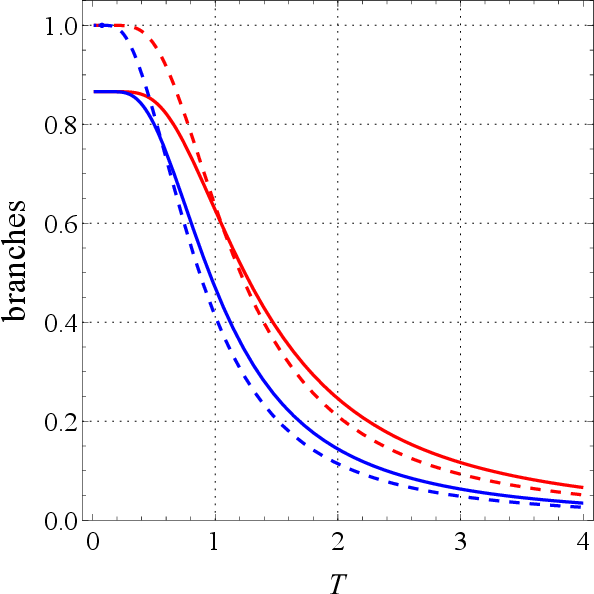,width=6cm}\put(-145,175){(a)}
\hspace{8mm}
\epsfig{file=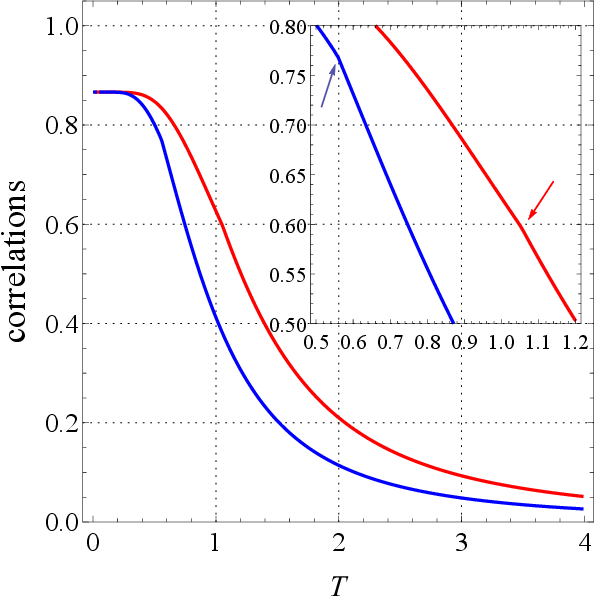,width=6cm}\put(-145,175){(b)}
\end{center}
\begin{center}
\caption{
Qubit-qutrit system with 
$B_1=0.3$, $B_2=-0.7$, $J=-1.4$, $J_z=1$, $K=0.2$, $K_1=-0.1$, $K_2=0.22$, $D_z=0.32$,
$\Gamma=-0.87$, and $\Lambda=0.31$.
(a)~Branches of quantum correlations
${\cal U}_0$ (blue solid line),
${\cal U}_1$ (blue dashed line),
${\cal F}_0$ (red solid line), and
${\cal F}_1$ (red dashed line)
as a function of $T$.
(b)~Quantum correlations $\cal U$ (blue line) and $\cal F$ (red line) versus  $T$. Arrow-up shows the abrupt change point for $\cal U$ and arrow-down indicates the similar point for $\cal F$.
}
\label{fig:3}
\end{center}
\end{figure}

Let us now put $J=-1.4$ while the other values are the same as before.
The behaviors of four branches ${\cal U}_0$, ${\cal U}_1$, ${\cal F}_0$,
and ${\cal F}_1$ are shown in Fig.~\ref{fig:3}(a). Besides, Fig.~\ref{fig:3}(b) depicts ${\cal U}$ 
and ${\cal F}$.
At high temperatures, LQU and LQFI are respectively branches ${\cal U}_1 (T)$ and ${\cal F}_1 (T)$, however, they experience sudden transitions as the system cools. Indeed, the intersections of 0- and 1-branches are seen clearly and they lead to abrupt changes in the behaviors of LQU and LQFI. At these points, the branches change from ${\cal U}_1 (T)$ and ${\cal F}_1 (T)$ to ${\cal U}_0 (T)$ and ${\cal F}_0 (T)$.
In the limit $T\to0$, these correlations reach the value $0.86609\ldots$, which is less
than 8/9.

%

%

\begin{figure}[!t]
\begin{center}
\epsfig{file=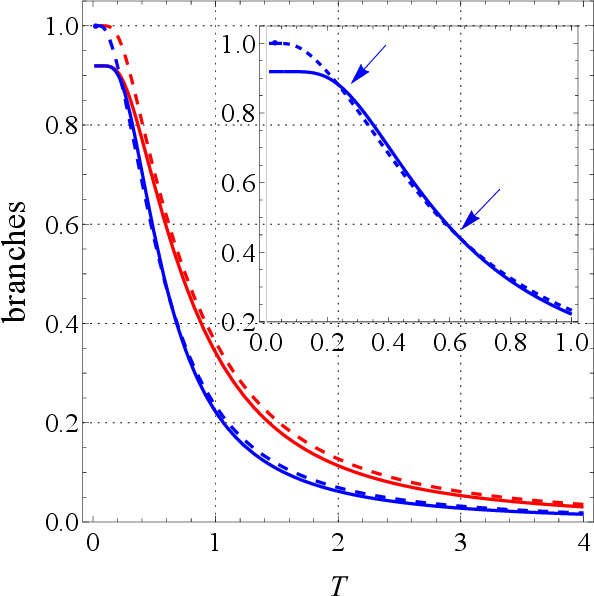,width=6cm}\put(-145,175){(a)}
\hspace{8mm}
\epsfig{file=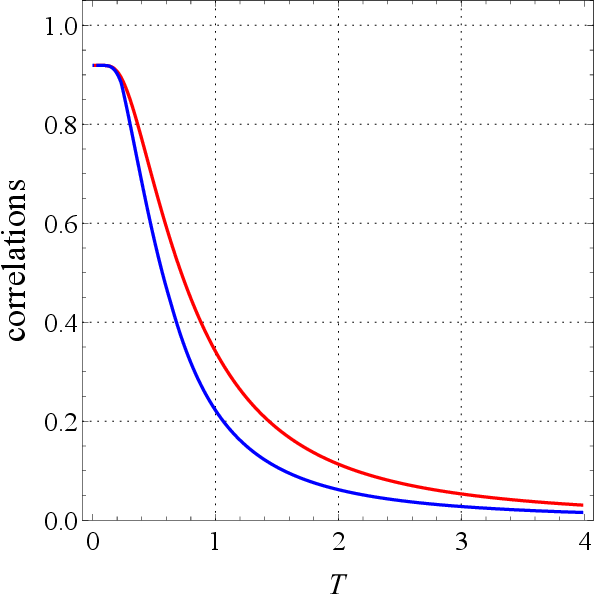,width=6cm}\put(-145,175){(b)}
\end{center}
\begin{center}
\caption{
Qubit-qutrit system with 
$B_1=0.7$, $B_2=0.3$, $J=-0.7$, $J_z=1$, $K=0.2$, $K_1=-0.1$, $K_2=0.22$, $D_z=0.32$,
$\Gamma=-0.87$, and $\Lambda=0.31$.
(a)~Branches of quantum correlations
${\cal U}_0$ (blue solid line),
${\cal U}_1$ (blue dashed line),
${\cal F}_0$ (red solid line), and
${\cal F}_1$ (red dashed line)
as a function of $T$.
(b)~Quantum correlations $\cal U$ (blue line) and $\cal F$ (red line) versus  $T$. Arrows-down show the abrupt change points for $\cal U$.
}
\label{fig:4}
\end{center}
\end{figure}
%

Fig. \ref{fig:4} illustrates the behaviors of four branches and two correlations with changed parameters $B_1=0.7$, $B_2=0.3$, $J=-0.7$ and the previously fixed parameters. We find another interesting phenomenon with these particular choices of system parameters, i.e. the existence of {\em more than one} sudden transition. From Fig. \ref{fig:4}(a), one can see that the branches ${\cal U}_0$ and
${\cal U}_1$ intersect twice, while branches ${\cal F}_0$ and
${\cal F}_1$ do not intersect. Therefore, as shown in Fig. \ref{fig:4}(b), the curve ${\cal U}(T)$ is piecewise-defined with two sudden changes, while the curve ${\cal F}(T)$ is smooth. Hence, the behaviors of both LQU and LQFI in our considered system with the same model parameters are different both quantitatively and qualitatively. Note that when $T\to 0$, these correlations reach the values $0.91905...$, which is more than $8/9$.

\begin{figure}[!t]
\begin{center}
\epsfig{file=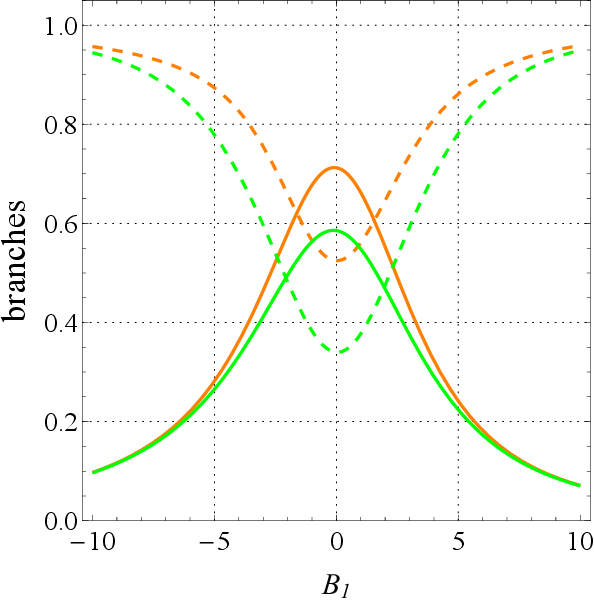,width=6cm}\put(-145,175){(a)}
\hspace{8mm}
\epsfig{file=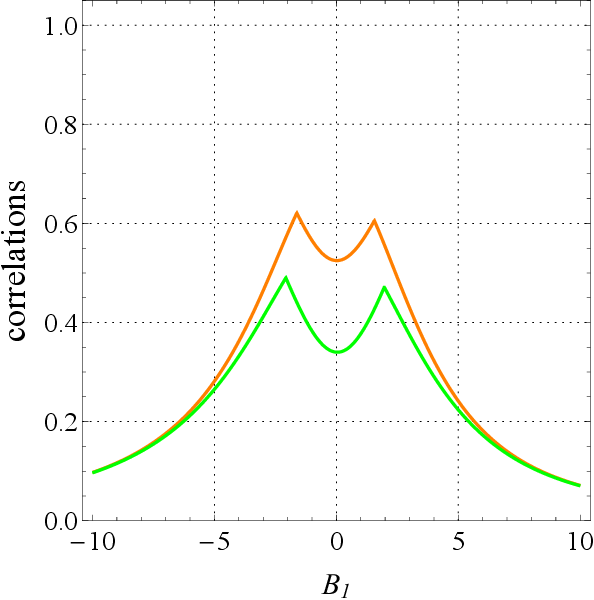,width=6cm}\put(-145,175){(b)}
\end{center}
\begin{center}
\caption{
Qubit-qutrit system with 
 $T=1$, $B_2=0$, $J=-2.5$, $J_z=-1$, $K=0.2$, $K_1=-0.1$, $K_2=0.22$, $D_z=0.32$,
$\Gamma=-0.87$, and $\Lambda=0.31$.
(a)~Branches of quantum correlations
${\cal U}_0$ (green solid line),
${\cal U}_1$ (green dashed line),
${\cal F}_0$ (orange solid line), and
${\cal F}_1$ (orange dashed line)
as a function of $B_1$.
(b)~Quantum correlations $\cal U$ (green line) and $\cal F$ (orange line) versus  $B_1$.
}
\label{fig:5}
\end{center}
\end{figure}
%

\subsubsection{Dependence on external fields}
As can be seen from Eq.~(\ref{eq:H}), the external magnetic field $B_1$ is applied to a qubit while the external magnetic field $B_2$ is applied to a qutrit. So, it is interesting to examine the effects of both $B_1$ and $B_2$ on quantum correlations.

\begin{figure}[!t]
\begin{center}
\epsfig{file=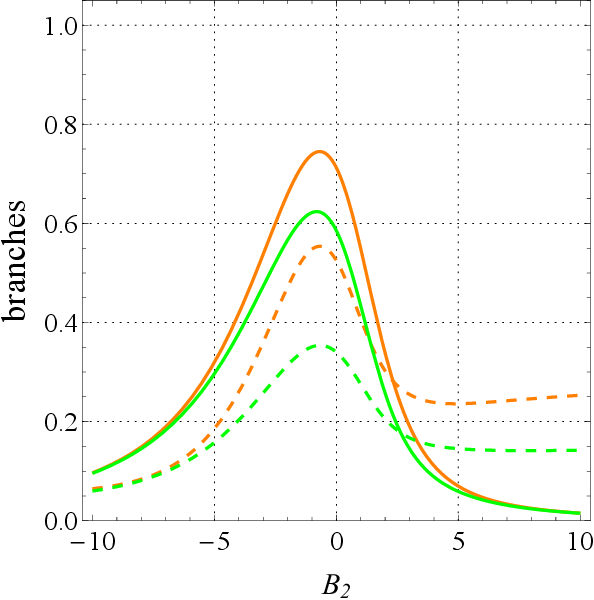,width=6cm}\put(-145,175){(a)}
\hspace{8mm}
\epsfig{file=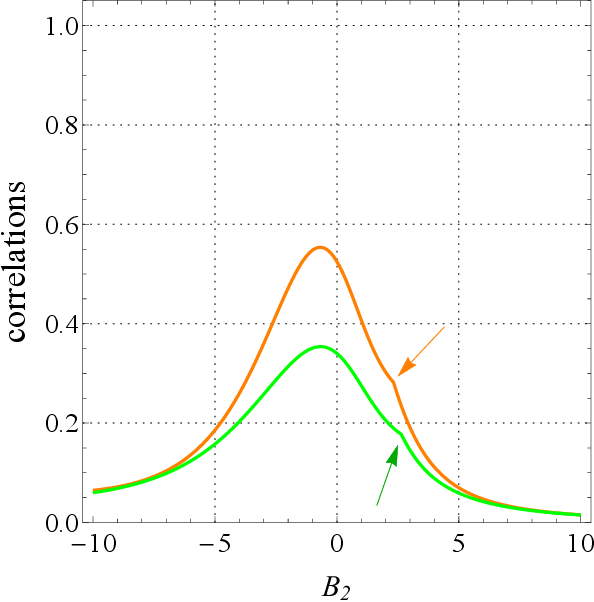,width=6cm}\put(-145,175){(b)}
\end{center}
\begin{center}
\caption{
Qubit-qutrit system with 
 $T=1$, $B_1=0$, $J=-2.5$, $J_z=-1$, $K=0.2$, $K_1=-0.1$, $K_2=0.22$, $D_z=0.32$,
$\Gamma=-0.87$, and $\Lambda=0.31$.
(a)~Branches of quantum correlations
${\cal U}_0$ (green solid line),
${\cal U}_1$ (green dashed line),
${\cal F}_0$ (orange solid line), and
${\cal F}_1$ (orange dashed line)
as a function of $B_2$.
(b)~Quantum correlations $\cal U$ (green line) and $\cal F$ (orange line) versus  $B_2$. Arrow-up shows the abrupt change point for $\cal U$ and arrow-down indicates the similar point for $\cal F$.
}
\label{fig:6}
\end{center}
\end{figure}

We plot the quantum correlations and their branches versus $B_1$ and $B_2$ with $B_2=0$ in Fig. \ref{fig:5} and $B_1=0$ in Fig. \ref{fig:6} when $J_z=-1$ (ferromagnetic case) at a fixed temperature $T=1$ with $J=-2.5$, $K=0.2$, $K_1=-0.1$, $K_2=0.22$, $D_z=0.32$,
$\Gamma=-0.87$, and $\Lambda=0.31$. 
Figure \ref{fig:5}(a) shows the behavior of branches related to LQU and LQFI, which intersect at certain points. As before, the reason for the sudden change of the correlations in Fig. \ref{fig:5}(b) is the intersections between the branches, whose roots are in Eqs. \eqref{eq:U} and \eqref{eq:F}. Indeed, we see two sharp changes in the behavior of both functions, represented by fractures on curves $\cal U$ and $\cal F$. 

Interestingly, the behavior of LQU and LQFI in the scenario where the external magnetic field is applied to a qutrit is significantly different from when it is applied to a qubit. Look at Fig. \ref{fig:6}; it shows the behavior of correlations and their branches against $B_2$. By comparing Figs. \ref{fig:5} and \ref{fig:6}, we notice that in addition to qualitative and quantitative differences between the behavior of functions, the quantum correlations experience only one sudden transition when the external magnetic field is applied to a qutrit. Thus, one can see obvious differences in the behavior of branches and correlations with respect to field $B_1$ acting on the qubit and field $B_2$ acting on the qutrit.

%

\begin{figure}[!t]
\begin{center}
\epsfig{file=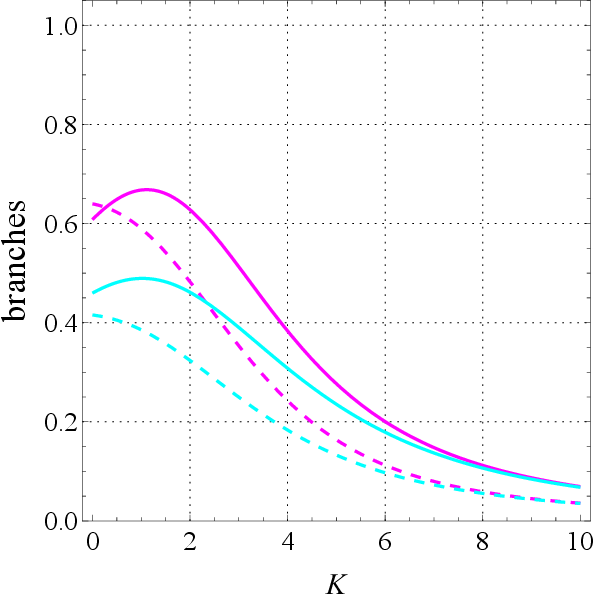,width=6cm}\put(-145,175){(a)}
\hspace{8mm}
\epsfig{file=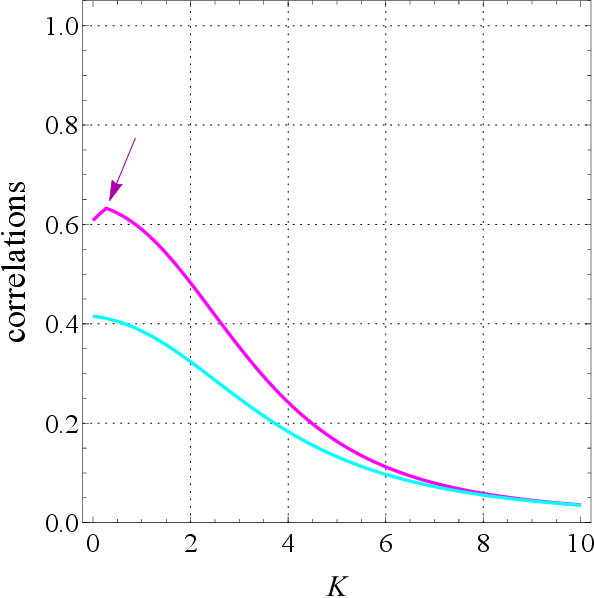,width=6cm}\put(-145,175){(b)}
\end{center}
\begin{center}
\caption{
Qubit-qutrit system with 
 $T=1$, $B_1=0.3$, $B_2=-0.7$ $J=-1.4$, $J_z=1$, $K_1=-0.1$, $K_2=0.22$, $D_z=0.32$,
$\Gamma=-0.87$, and $\Lambda=0.31$.
(a)~Branches of quantum correlations
${\cal U}_0$ (cyan solid line),
${\cal U}_1$ (cyan dashed line),
${\cal F}_0$ (magenta solid line), and
${\cal F}_1$ (magenta dashed line)
as a function of $K$.
(b)~Quantum correlations $\cal U$ (cyan line) and $\cal F$ (magenta line) versus  $K$. Arrow-down shows the abrupt change point for $\cal F$.
}
\label{fig:7}
\end{center}
\end{figure}
%

\begin{figure}[!t]
\begin{center}
\epsfig{file=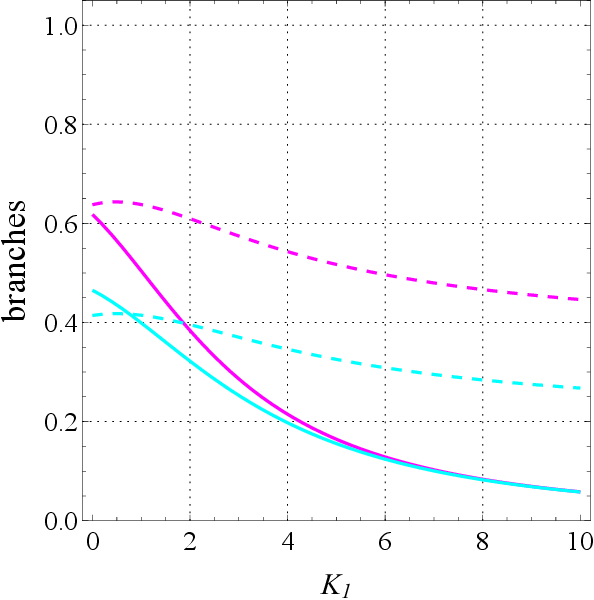,width=6cm}\put(-145,175){(a)}
\hspace{8mm}
\epsfig{file=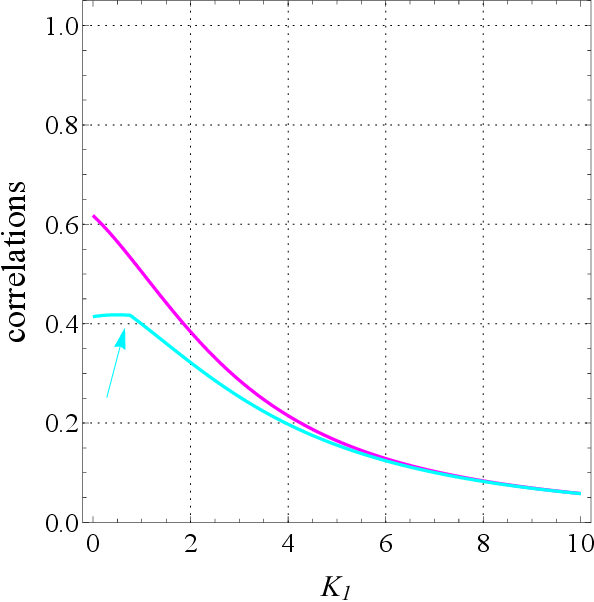,width=6cm}\put(-145,175){(b)}
\end{center}
\begin{center}
\caption{
Qubit-qutrit system with 
 $T=1$, $B_1=0.3$, $B_2=-0.7$ $J=-1.4$, $J_z=1$, $K=0.2$, $K_2=0.22$, $D_z=0.32$,
$\Gamma=-0.87$, and $\Lambda=0.31$.
(a)~Branches of quantum correlations
${\cal U}_0$ (cyan solid line),
${\cal U}_1$ (cyan dashed line),
${\cal F}_0$ (magenta solid line), and
${\cal F}_1$ (magenta dashed line)
as a function of $K_1$.
(b)~Quantum correlations $\cal U$ (cyan line) and $\cal F$ (magenta line) versus  $K_1$. Arrow-up shows the abrupt change point for $\cal U$.
}
\label{fig:8}
\end{center}
\end{figure}
%
\begin{figure}[!t]
\begin{center}
\epsfig{file=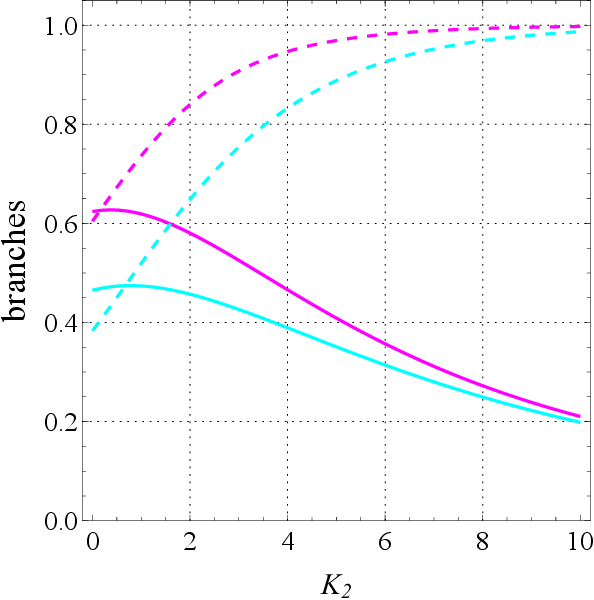,width=6cm}\put(-145,175){(a)}
\hspace{8mm}
\epsfig{file=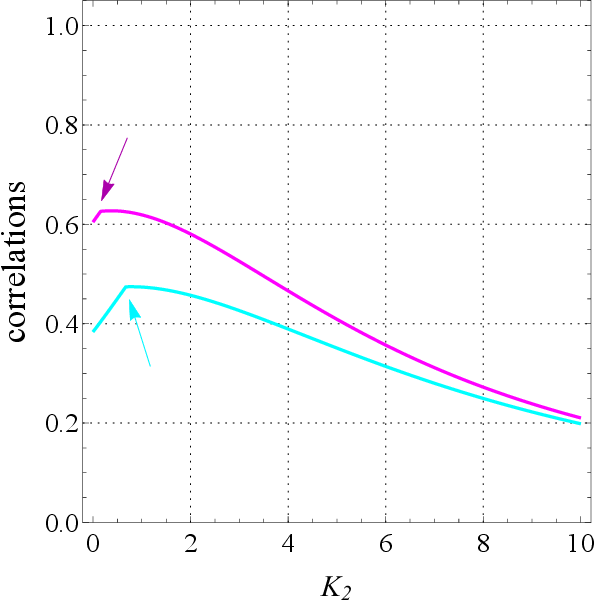,width=6cm}\put(-145,175){(b)}
\end{center}
\begin{center}
\caption{
Qubit-qutrit system with 
 $T=1$, $B_1=0.3$, $B_2=-0.7$ $J=-1.4$, $J_z=1$, $K=0.2$, $K_1=-0.1$, $D_z=0.32$,
$\Gamma=-0.87$, and $\Lambda=0.31$.
(a)~Branches of quantum correlations
${\cal U}_0$ (cyan solid line),
${\cal U}_1$ (cyan dashed line),
${\cal F}_0$ (magenta solid line), and
${\cal F}_1$ (magenta dashed line)
as a function of $K_2$.
(b)~Quantum correlations $\cal U$ (cyan line) and $\cal F$ (magenta line) versus  $K_2$. Arrow-up shows the abrupt change point for $\cal U$ and arrow-down indicates the similar point for $\cal F$.
}
\label{fig:9}
\end{center}
\end{figure}
%

\subsubsection{Dependence on uniaxial and planar anisotropies}
Since the uniaxial and planar one-ion anisotropies, respectively $K$ and $K_1$, only affect the anisotropy of qutrit while the uniaxial two-ion anisotropy $K_2$ affects the anisotropy of both qubit and qutrit, it can be interesting to investigate the behavior of quantum correlations when these parameters are changed.

By keeping some parameters of the system constant, in Figs. \ref{fig:7} to \ref{fig:9}, we have drawn the branches and correlations in terms of $K$, $K_1$ and $K_2$. Figure \ref{fig:7} shows that the behavior of LQU is smooth because no intersection between branches ${\cal U}_0$ and ${\cal U}_1$  is observed. However, a sudden change in the behavior of LQFI is observed, which originates from the intersection of branches ${\cal F}_0$ and ${\cal F}_1$.

Significantly, the scenario changes dramatically if we plot branches and correlations versus $K_1$ (look at Fig. \ref{fig:8}), meaning that the sudden transition is only visible in the behavior of LQU while the behavior of LQFI is monotonic. 

We discover another interesting result when we plot the branches and correlations as a function of $K_2$, as seen from Fig. \ref{fig:9}. Both LQU and LQFI experience one sudden transition in specific values of the qubit and qutrit uniaxial anisotropy.

Thereby, the correlations and their branches exhibit different properties when different anisotropies are considered for the system. We also emphasize our previous statement that the behavior of correlations differs not only quantitatively but also qualitatively.

\section{
Conclusion
}
\label{subsubsec:Concl}
We have derived the compact closed forms of two discord-like quantum correlation measures (LQU and LQFI) for general qubit-qutrit AS states. As an application of the derived formulas for LQU and LQFI, we explored the behavior of quantum correlations at thermal equilibrium in axisymmetric hybrid qubit-qutrit systems. For that, we considered the most general Hamiltonian of the qubit-qutrit system commuting with $z$-component of total spin ${\cal S}_z=s_z\otimes{\rm I_3}+{\rm I_2}\otimes S_z$. New features were observed in the behavior of quantum correlations, which are significant. For instance, sudden changes in the behavior of quantum correlations were found with a smooth change in temperature or other parameters. Besides, it was revealed that there may be a series (sequence) of such changes for certain choices of system parameters.
Interestingly, in some cases, sudden transitions were observed in the behavior of LQU, but not in the behavior of LQFI, and vice versa. Thus, LQU and LQFI exhibit different properties when certain choices of parameters are considered for the system. Hence, in our system with the same parameters, the behaviors of LQU and LQFI can distinguish not only quantitatively but also qualitatively.

We believe that our derived formulas for LQU and LQFI can also be beneficial in other interesting problems with general qubit-qutrit AS states, for example, when investigating the environmental effects on quantum correlations in open systems.

\section*{Data availability}
All data generated or analysed during this study are included in this published article.
\\
\\
\section*{Acknowledgements}
M.A.Y. was supported in part by a state task, the state registration number of the Russian Federation is \#124013000760-0. S.H. and M.G. were supported by Semnan University under Contract No. 21270.
\\
\\
\section*{Author contributions}
M.A.Y. and S.H. have contributed to writing the manuscript and interpreting the results. Thorough checking of the manuscript was done by M.A.Y. and M.G. The final draft of the manuscript was revised by all authors.
\\
\\
\section*{Disclosures}
The authors declare that they have no known competing financial interests.


\section{Methods}
\label{sec:Model}
\textbf{Density matrix and its diagonalization.} The axial symmetry group $U(1)$, consisting of rotations $R_z(\phi)=\exp(-i\phi{\cal S}_z)$ around the $z$-axis on angles $\phi$,
means for the density matrix $\rho$
 of a qubit-qutrit system 
that it commutes with the
$z$-component of total spin
${\cal S}_z=s_z\otimes{\rm I_3}+{\rm I_2}\otimes S_z={\rm diag}\,[3/2,1/2,-1/2,1/2,-1/2,-3/2]$.
The most general Hermitian matrix which commutes with ${\cal S}_z$ has the form
\begin{equation}
   \label{eq:rho}
   \rho=
	 \left(
      \begin{array}{cllccc}
      p_1&\ &\ &\ &\ \\
      \ &a&0&u&0&\ \\
			\ &0&b&0&v&\ \\
      \ &u^*&0&c&0&\ \\
      \ &0&v^*&0&d&\ \\
			\ &\ &\ &\ &\ &p_6
      \end{array}
   \right).
\end{equation}
This matrix is quasi-diagonal
$(1\times1)-(4\times4)-(1\times1)$,
where $4\times4$-subblock is sparse and has ``checkerboard'' structure.
Remarkably, the set of matrices of such a type are algebraically closed: their
sums and products preserve the same form.
These matrices play the same important role for the qubit-qutrit models as X-matrices
for the two-qubit
models and therefore we will shortly call them the AS-matrices.

In general, such a nonnormalized  Hermitian six-by-six AS-matrix depends on ten real
independent parameters. 
However taken into account normalization condition ${\rm Tr}\rho=1$ (i.e.,
$p_1+p_6+a+b+c+d=1$), the number of independent parameters for the density matrix is
reduced to nine.

It is clear that after the permutation of 3-rd and 4-th rows and columns of AS matrix
(\ref{eq:rho}),
its $4\times4$ inner subblock splits into two $2\times2$ subblocks:
\begin{equation}
   \label{eq:rho1}
   \rho^\prime=P_{34}\rho P_{34}^t=
	 \left(
      \begin{array}{cllccc}
      p_1&\ &\ &\ &\ \\
      \ &a&u&\ &\ &\ \\
			\ &u^*&c&\ &\ &\ \\
      \ &\ &\ &b&v&\ \\
      \ &\ &\ &v^*&d&\ \\
			\ &\ &\ &\ &\ &p_6
      \end{array}
   \right),
\end{equation}
where $P_{34}$ is the  orthogonal transformation
\begin{equation}
   \label{eq:P34}
   P_{34}=
	 \left(
      \begin{array}{cccccc}
      1&.&.&.&.&.\\
      .&1&.&.&.&.\\
      .&.&.&1&.&.\\
      .&.&1&.&.&.\\
      .&.&.&.&1&.\\
      .&.&.&.&.&1
      \end{array}
   \right)=P_{34}^t
\end{equation}
(the superscript $t$ stands for matrix transpose).
Due to a nonnegativity of any density matrix, it follows that $p_1,a,b,c,d,p_6\ge0$,
$ac\ge|u|^2$ and $bd\ge |v|^2$.

The eigenvalues of the density matrix (\ref{eq:rho}) are equal to $p_1$, $p_6$ and
\begin{equation}
   \label{eq:p_2-5}
   p_{2,3}=\frac{1}{2}\Big(a+c\pm\sqrt{(a-c)^2+4|u|^2}\Big),\quad
   p_{4,5}=\frac{1}{2}\Big(b+d\pm\sqrt{(b-d)^2+4|v|^2}\Big).
\end{equation}
To fully 
diagonalize the density matrix (\ref{eq:rho}), we now use
the unitary transformation $R$ which is built by eigenvectors of the density
matrix (\ref{eq:rho1}), i.e. from vectors of $2\times2$ subblocks
(in this connection see, e.g., \cite{M57})
\begin{equation}
   \label{eq:R}
   R=
	 \left(
      \begin{array}{cccccc}
      1&.&.&.&.&.\\
      .&q_1/\sqrt{q_1^2+|u|^2}&u/\sqrt{q_1^2+|u|^2}&.&.&.\\
      .&u^*/\sqrt{q_1^2+|u|^2}&-q_1/\sqrt{q_1^2+|u|^2}&.&.&.\\
      .&.&.&q_2/\sqrt{q_2^2+|v|^2}&v/\sqrt{q_2^2+|v|^2}&.\\
      .&.&.&v^*/\sqrt{q_2^2+|v|^2}&-q_2/\sqrt{q_2^2+|v|^2}&.\\
      .&.&.&.&.&1\\
      \end{array}
   \right)=R^\dagger,
\end{equation}
where
\begin{equation}
   \label{eq:q1}
   q_1=\frac{1}{2}\Big(a-c+\sqrt{(a-c)^2+4|u|^2}\Big),\quad
   q_2=\frac{1}{2}\Big(b-d+\sqrt{(b-d)^2+4|v|^2}\Big);
\end{equation}
$R^\dagger=R$, $R^\dagger R=1$ and
$RP_{34}\rho P_{34}R={\rm diag}[p_1,p_2,p_3,p_4,p_5,p_6]$.
Note useful expressions:
\begin{equation}
   \label{eq:q2a}
   q_1^2=(a-c)q_1+|u|^2,\quad
   q_2^2=(b-d)q_2+|v|^2.
\end{equation}
Here, one should keep the carefulness when $|u|$ or $|v|$ equals zero.

Using transformations (\ref{eq:P34}) and (\ref{eq:R}), we now get local spin matrices
$\sigma_\mu\otimes I_3$ ($\mu=x,y,z$) in the diagonal representation of the density
matrix (i.e., we find the sets of matrix elements
$\langle m|\sigma_\mu\otimes I_3|n\rangle$)
\begin{eqnarray}
   \label{eq:xI3}
   &&RP_{34}(\sigma_x\otimes I_3)P_{34}R=
   \nonumber\\
	 &&\left(
      \begin{array}{cccccc}
      .&\frac{u^*}{\sqrt{q_1^2+|u|^2}}&\frac{-q_1}{\sqrt{q_1^2+|u|^2}}&.&.&.\\
      \frac{u}{\sqrt{q_1^2+|u|^2}}&.&.&\frac{q_1v^*}{\sqrt{(q_1^2+|u|^2)(q_2^2+|v|^2)}}&\frac{-q_1q_2}{\sqrt{(q_1^2+|u|^2)(q_2^2+|v|^2)}}&.\\ \\
      \frac{-q_1}{\sqrt{q_1^2+|u|^2}}&.&.&\frac{u^*v^*}{\sqrt{(q_1^2+|u|^2)(q_2^2+|v|^2)}}&\frac{-q_2u^*}{\sqrt{(q_1^2+|u|^2)(q_2^2+|v|^2)}}&.\\
			.&\frac{q_1v}{\sqrt{(q_1^2+|u|^2)(q_2^2+|v|^2)}}&\frac{u v}{\sqrt{(q_1^2+|u|^2)(q_2^2+|v|^2)}}&.&.&\frac{q_2}{\sqrt{q_2^2+|v|^2}}\\ \\
			.&\frac{-q_1q_2}{\sqrt{(q_1^2+|u|^2)(q_2^2+|v|^2)}}&\frac{-q_2u}{\sqrt{(q_1^2+|u|^2)(q_2^2+|v|^2)}}&.&.&\frac{v^*}{\sqrt{q_2^2+|v|^2}}\\ \\
      .&.&.&\frac{q_2}{\sqrt{q_2^2+|v|^2}}&\frac{v}{\sqrt{q_2^2+|v|^2}}&.\\
      \end{array}
   \right),
   \nonumber\\
\end{eqnarray}
\begin{eqnarray}
   \label{eq:yI3}
   &&RP_{34}(\sigma_y\otimes I_3)P_{34}R=
   \nonumber\\
	 &&\left(
      \begin{array}{cccccc}
      .&\frac{-i u^*}{\sqrt{q_1^2+|u|^2}}&\frac{iq_1}{\sqrt{q_1^2+|u|^2}}&.&.&.\\
      \frac{iu}{\sqrt{q_1^2+|u|^2}}&.&.&\frac{-iq_1v^*}{\sqrt{(q_1^2+|u|^2)(q_2^2+|v|^2)}}&\frac{iq_1q_2}{\sqrt{(q_1^2+|u|^2)(q_2^2+|v|^2)}}&.\\ \\
      \frac{-iq_1}{\sqrt{q_1+|u|^2}}&.&.&\frac{-iu^*v^*}{\sqrt{(q_1^2+|u|^2)(q_2^2+|v|^2)}}&\frac{iq_2u^*}{\sqrt{(q_1^2+|u|^2)(q_2^2+|v|^2)}}&.\\
			.&\frac{iq_1v}{\sqrt{(q_1^2+|u|^2)(q_2^2+|v|^2)}}&\frac{iuv}{\sqrt{(q_1^2+|u|^2)(q_2^2+|v|^2)}}&.&.&\frac{-iq_2}{\sqrt{q_2^2+|v|^2}}\\ \\
			.&\frac{-iq_1q_2}{\sqrt{(q_1^2+|u|^2)(q_2^2+|v|^2)}}&\frac{-iq_2u}{\sqrt{(q_1^2+|u|^2)(q_2^2+|v|^2)}}&.&.&\frac{-i v^*}{\sqrt{q_2^2+|v|^2}}\\ \\
      .&.&.&\frac{iq_2}{\sqrt{q_2^2+|v|^2}}&\frac{iv}{\sqrt{q_2^2+|v|^2}}&.\\
      \end{array}
   \right)
   \nonumber\\
\end{eqnarray}
and
\begin{equation}
   \label{eq:zI3}
   RP_{34}(\sigma_z\otimes I_3)P_{34}R=
	 \left(
      \begin{array}{cccccc}
      1&.&.&.&.&.\\
      .&\frac{q_1^2-|u|^2}{q_1^2+|u|^2}&\frac{2q_1u}{q_1^2+|u|^2}&.&.&.\\ \\
      .&\frac{2q_1u^*}{q_1^2+|u|^2}&-\frac{q_1^2-|u|^2}{q_1^2+|u|^2}&.&.&.\\
      .&.&.&\frac{q_2^2-|v|^2}{q_2^2+|v|^2}&\frac{2q_2 v}{q_2^2+|v|^2}&.\\ \\
      .&.&.&\frac{2q_2v^*}{q_2^2+|v|^2}&-\frac{q_2^2-|v|^2}{q_2^2+|v|^2}&.\\
      .&.&.&.&.&-1\\
      \end{array}
   \right).
\end{equation}

\vspace{2cm}
\section*{References}



\end{document}